\pdfoutput=1  
\documentclass[10pt,conference]{IEEEtran}
\IEEEoverridecommandlockouts

\usepackage{cite}
\usepackage{amsmath,amssymb,amsfonts}
\usepackage{algorithmic}
\usepackage{graphicx}
\usepackage{textcomp}
\usepackage{xcolor}
\usepackage{comment}
\usepackage[acronym]{glossaries}
\usepackage[free-standing-units,per-mode=repeated-symbol,binary-units=true,detect-weight=true,detect-family=true]{siunitx}[=v2]
\usepackage{lscape}
\usepackage{multirow}

\usepackage{threeparttable} 
\usepackage{booktabs}  
\usepackage{tikz}

\def\BibTeX{{\rm B\kern-.05em{\sc i\kern-.025em b}\kern-.08em
    T\kern-.1667em\lower.7ex\hbox{E}\kern-.125emX}}

\usepackage{subcaption}
\begin{document}

\title{Design in Tiles: Automating GEMM Deployment on Tile-Based Many-PE Accelerators}
\ifx\blind\undefined
    \author{
    \IEEEauthorblockN{Aofeng Shen\textsuperscript{1}, Chi Zhang\textsuperscript{1}, Yakup Budanaz\textsuperscript{2}, Alexandru Calotoiu\textsuperscript{2}, Bowen Wang\textsuperscript{1}, Torsten Hoefler\textsuperscript{2}, and Luca Benini\textsuperscript{1}}
    \IEEEauthorblockA{ETH Z\"{u}rich, Z\"{u}rich, Switzerland\\
    \textsuperscript{1}\{aoshen, chizhang, bowwang, lbenini\}@iis.ee.ethz.ch,\quad
    \textsuperscript{2}\{yakupkoray.budanaz, alexandru.calotoiu, torsten.hoefler\}@inf.ethz.ch}
    }
\else
    \author{\centering{\textit{Authors omitted for blind review.}\vspace{28pt}}}
\fi

\newacronym{hbm}{HBM}{High Bandwidth Memory}
\newacronym{llm}{LLM}{Large Language Model}
\newacronym{tdp}{TDP}{Thermal Design Power}
\newacronym{dnn}{DNN}{Deep Neural Network}
\newacronym{mha}{MHA}{Multi-Head Attention}
\newacronym{ml}{ML}{Machine Learning}
\newacronym{pe}{PE}{Processing Element}
\newacronym{ce}{CE}{Compute Element}
\newacronym{soa}{SoA}{state-of-the-art}
\newacronym{dma}{DMA}{Direct Memory Access}
\newacronym{noc}{NoC}{Network on Chip}
\newacronym{fu}{FU}{Function Unit}
\newacronym{ai}{AI}{Artificial Intelligence}
\newacronym{rvv}{RVV}{RISC-V Vector}
\newacronym{ge}{GE}{Gate Equivalent}
\newacronym{FA}{FA}{FlashAttention}
\newacronym{gemm}{GEMM}{General Matrix Multiplication}
\newacronym{dit}{DiT}{Design in Tiles}
\newacronym{ir}{IR}{Intermediate Representation}
\newacronym{hpc}{HPC}{High Performance Computing}
\newacronym{spm}{SPM}{Scratchpad Memory}
\newacronym{dace}{DaCe}{Data-Centric Parallel Programming}
\newacronym{mmad}{MMAD}{Matrix Multiplication Addition}
\newacronym{bsp}{BSP}{Bulk Synchronous Parallel}
\newacronym{ast}{AST}{Abstract Syntax Tree}
\newacronym{sdfg}{SDFG}{Stateful Dataflow Multigraph}
\newacronym{ffn}{FFN}{Feed Forward Network}

\maketitle

\begin{abstract}
Tile-based many-\gls{pe} accelerators can achieve competitive performance on \gls{gemm}, but they are hard to program, as their optimal software mapping is deeply coupled with hardware architecture. We propose "\gls{dit}", an automated framework connecting a deployment toolchain with a configurable executable model for these accelerators. For evaluation, we apply our framework to \gls{gemm} targeting a large acceleration configuration (e.g., 32$\times$32 tiles, 1979 TFLOPS@FP8, 4 TB/s Bandwidth) comparable to an NVIDIA GH200. We achieve higher PE utilization than GH200's expert-tuned \gls{gemm} libraries, achieving 1.2--2.0$\times$ speedup across diverse matrix shapes.
\end{abstract}


\section{Introduction}

\begin{figure}[b]
    \centering
    \begin{subfigure}[b]{0.52\linewidth}
        \centering
        \includegraphics[width=0.88\linewidth]{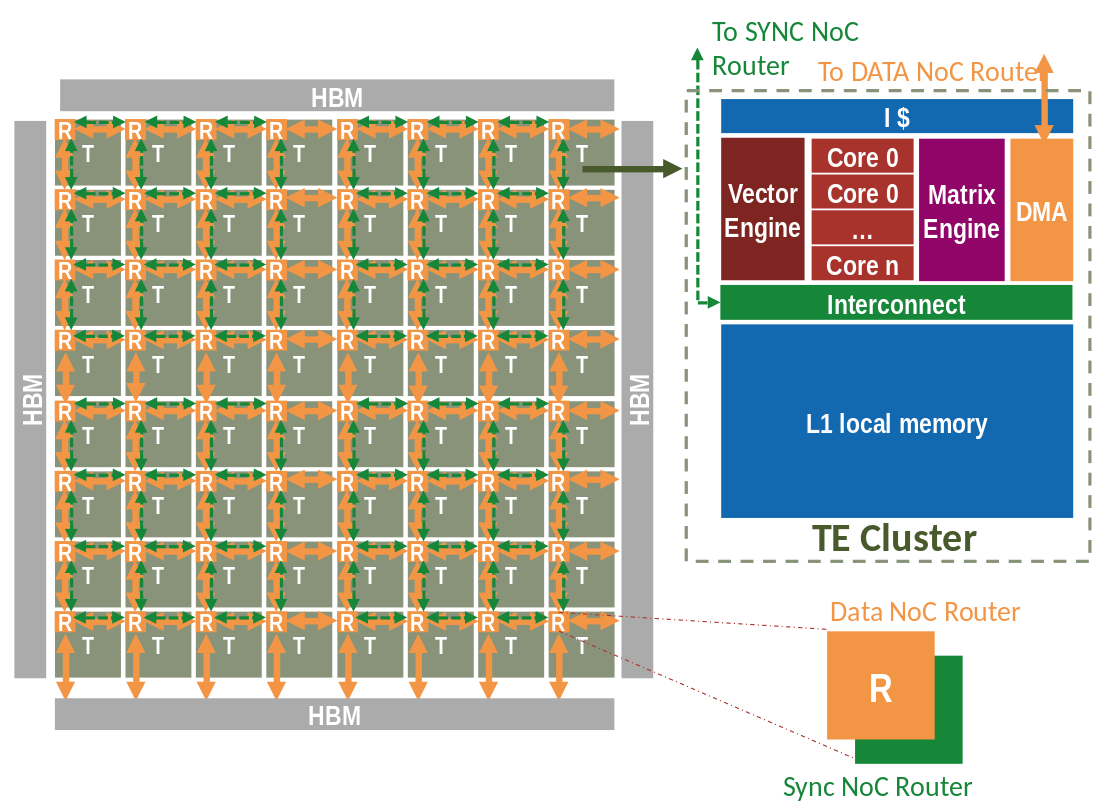}
        \caption{SoftHier architecture}
        \label{fig:softhier}
    \end{subfigure}\hfill
    \begin{subfigure}[b]{0.46\linewidth}
        \centering
        \includegraphics[width=\linewidth]{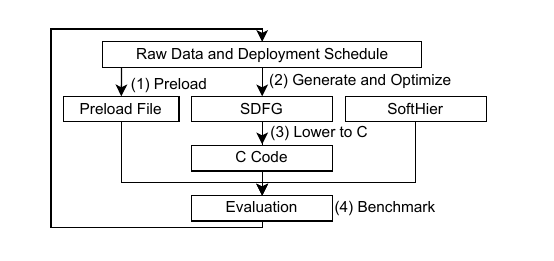}
        \caption{DiT workflow}
        \label{fig:workflow}
    \end{subfigure}

    \vspace{2pt}
    \begin{subfigure}[b]{\linewidth}
        \centering
        \includegraphics[width=0.2\linewidth, clip, trim=20pt 20pt 20pt 20pt]{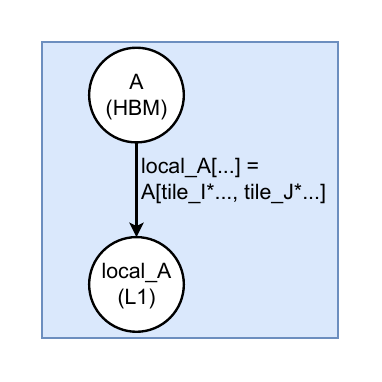}\hfill
        \includegraphics[width=0.2\linewidth, clip, trim=20pt 20pt 20pt 20pt]{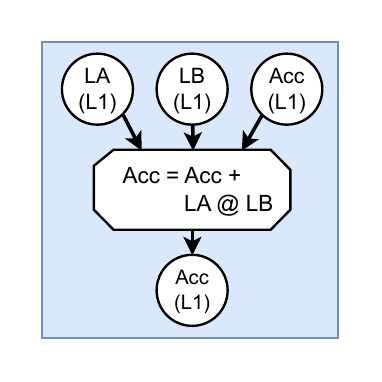}\hfill
        \includegraphics[width=0.2\linewidth, clip, trim=20pt 20pt 20pt 20pt]{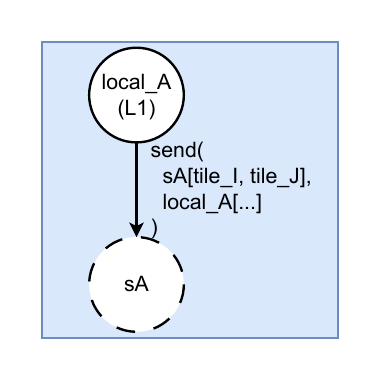}\hfill
        \includegraphics[width=0.2\linewidth, clip, trim=20pt 20pt 20pt 20pt]{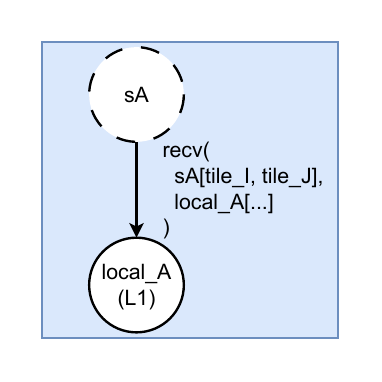}
        \caption{SDFG primitives}
        \label{fig:sdfg_example}
    \end{subfigure}
    \vspace{-3pt}
    \caption{\gls{dit} framework: (a)~SoftHier, (b)~workflow, (c)~\gls{sdfg} primitives.}
    \label{fig:framework}
    \vspace{-14pt}
\end{figure}

Scaling traditional GPUs for large \gls{ai} and \gls{hpc} workloads reveals a utilization problem rooted in their hardware-managed cache hierarchy: even with expert-tuned CUTLASS, the newer, larger GH200 shows \emph{lower} average \gls{gemm} utilization than the older A100. Tile-based many-\gls{pe} accelerators address this (e.g., Tesla Dojo, Tenstorrent Blackhole): they replace unified caches with compute tiles, each pairing a software-managed L1 \gls{spm} with local \glspl{pe}, and connect these tiles through a programmable \gls{noc} to distributed \gls{hbm}, giving direct control over the on-chip dataflow. A recent study~\cite{chi2025} shows that hardware-based \gls{noc} collective communication lets such accelerators outperform GPUs on \gls{mha}.

However, they demand software-managed mapping, scheduling, and orchestration for hundreds of tiles, whose optimal strategy depends strongly on the hardware, making manual re-tuning impractical. Prior tile-based schedule-search work either lacks detailed \gls{gemm} deployment or relies on fixed communication patterns, none exploiting \gls{noc} collectives. We close this gap with \gls{dit}, an automated \gls{gemm} deployment framework for tile-based accelerators, contributing: an end-to-end flow (code generation, compilation, cycle-accurate analysis, verification); an \gls{ir} modeling per-PE workload, data movement, and inter-tile communication; a parameterized schedule abstraction (tiling/mapping, distributed-HBM layouts, collective patterns); and a \gls{gemm} study with a 1.2--2.0$\times$ speedup over GH200's expert-tuned libraries.

\section{Design Overview}

\begin{figure*}[t]
    \centering
    \includegraphics[width=\textwidth]{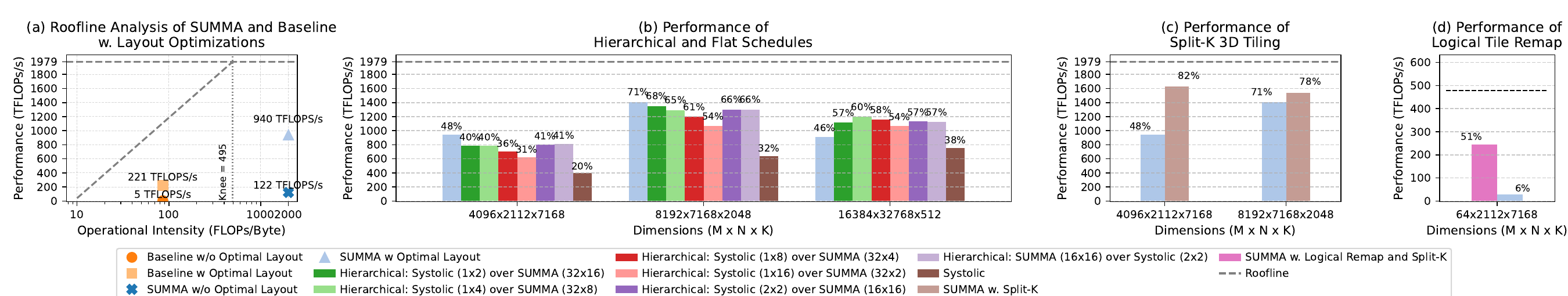}
    \vspace{-16pt}
    \caption{Optimization breakdown: (a) roofline of baseline and SUMMA with/without layout optimization; (b) hierarchical vs.\ flat 2D schedules; (c) 2D vs.\ 3D split-K tiling; (d) logical cluster-dimension remapping for flat GEMM.}
    \label{fig:eval_breakdown}
    \vspace{-12pt}
\end{figure*}

Our framework couples an executable model (SoftHier) with a deployment tool (DaCe) in an end-to-end workflow.

\textbf{SoftHier} is an RTL-calibrated ($<$10\% timing inaccuracy) simulator for tile-based many-PE accelerators, built on GVSoC and faithfully modeling \gls{noc}, \gls{dma}, and memory-controller contention, with \gls{hbm} modeled via DRAMSys5 for realistic bank-level timing. As shown in Figure~\ref{fig:softhier}, each tile-engine (TE) cluster holds \glspl{pe} (scalar, vector, matrix engine), \glspl{dma}, and a local L1 \gls{spm}, connected by data and sync \glspl{noc} with off-chip \gls{hbm} at the grid edges. Its key feature is hardware-supported collective communication: a mask-based scheme lets one tile broadcast to, or reduce from, a custom group of tiles via coordinate masks, without software fan-out.

\textbf{DaCe}~\cite{dace} is our deployment tool, built around the data-centric \gls{sdfg} \gls{ir} that captures all tile data movement. As shown in Figure~\ref{fig:sdfg_example}, data movement is a path between access nodes, computation an \gls{mmad} tasklet, and inter-tile transfers pass through per-tile stream arrays (send/receive). These primitives compose into dataflow schedules tailored to different GEMM workloads (Section~\ref{sec:abstraction}). Our SoftHier backend lowers the \gls{sdfg} to C, compiled to RISC-V. The workflow (Figure~\ref{fig:workflow}) turns a high-level schedule and layout into a verified executable: preload tensors across HBM channels, generate and optimize the \gls{sdfg}, lower to a binary, and benchmark against reference outputs.

\section{Tile-based Deployment Schedule Abstraction}
\label{sec:abstraction}
DiT describes how a workload is decomposed and mapped onto SoftHier through three components, enabling automated code generation from a high-level description.

\textbf{Tiling and Mapping} define each tile's computation under an output-stationary strategy: one tile computes a whole output tile (2D GEMM), or several collaborate with a reduction (3D split-K GEMM). Since the physical grid is fixed while the optimal mapping depends on the shape, a \emph{cluster-index remapping} reinterprets it as a logical grid (e.g., $1\times16$), automatically generating the matching physical-grid masks for its collectives.

\textbf{Data Layout} explicitly distributes data across SoftHier's multi-channel \gls{hbm} to avoid channel contention: a \emph{split scheme} partitions the matrix into blocks and a \emph{placement scheme} maps a block's tiles into a channel's address space to localize accesses.

\textbf{Dataflow} specifies movement across the hierarchy (\gls{hbm}$\leftrightarrow$L1 and tile-to-tile over the \gls{noc}) with communication/computation overlap via double buffering. Using mask-based collectives, DiT supports arbitrary reduction patterns; the implemented primitives are \textbf{SUMMA}, \textbf{Systolic}, \textbf{Hierarchical}, and \textbf{Split-K} (Figure~\ref{fig:eval_breakdown}), each a \gls{bsp} superstep (computation, communication, barrier) in a Python \gls{ast}.

\textbf{Automated Exploration.} Deployment is a search over schedules combining these components, subject to hardware constraints (L1 capacity, matrix-engine-friendly tiles, single-NoC-dimension access). DiT enumerates valid combinations, prunes infeasible ones, and simulates each on SoftHier, auto-selecting the best (300--1000+ candidates per shape, parallelized across CPU nodes).

\section{Evaluation}
We evaluate \gls{dit} on SoftHier spec-matched to commercial GPUs, reporting the best layout candidate. The main setup matches NVIDIA GH200 peak (a $32\times32$ cluster of $64\times16$ matrix engines, 384~KB L1 each, 1979~TFLOPS@FP8, 4~TB/s).

\textbf{Optimization breakdown.} Figure~\ref{fig:eval_breakdown} quantifies each optimization: \emph{(a)}~an optimized layout lifts the memory-bound baseline from 5 to 221~TFLOPS and a SUMMA dataflow raises operational intensity to 940~TFLOPS; \emph{(b)}~collective multicast beats the systolic pattern, though excess pipeline stages hurt store-intensive cases; \emph{(c)}~for irregular shapes, 3D split-K tiling with NoC reduction reaches 78--82\% utilization vs.\ $\sim$48\% for 2D; and \emph{(d)}~for flat GEMM, cluster-dimension remapping ($32\times32\to1\times1024$) lifts utilization from 6\% to 51\%.

\begin{figure}[h!]
    \centering
    \includegraphics[width=0.78\linewidth]{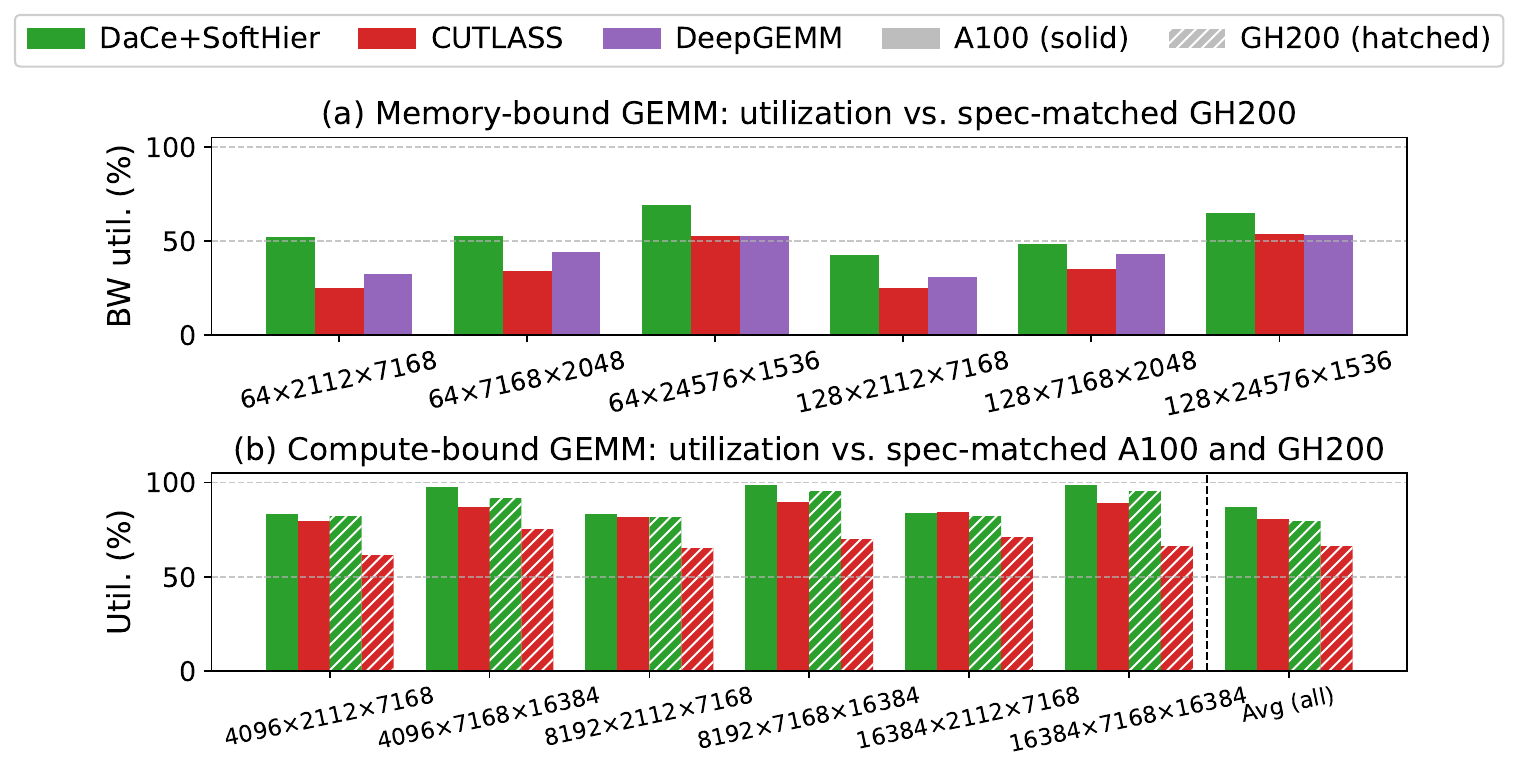}
    \vspace{-6pt}
    \caption{Spec-matched comparison with GPUs: (a)~memory-bound and (b)~compute-bound GEMM utilization, sharing one legend.}
    \label{fig:portability}
    \vspace{-10pt}
\end{figure}

\textbf{Spec-matched Comparison with GPUs.} We spec-match SoftHier instances to the A100 ($312$~TFLOPS@FP16) and GH200 and compare \gls{dit} against their expert-tuned libraries (CUTLASS, DeepGEMM) on the corresponding real GPUs, picking the best kernel per shape. On frequently used DeepSeek-V3 shapes, \gls{dit} achieves $1.2$--$1.5\times$ higher TFLOPS on compute-bound GEMM and, on flat memory-bound GEMM, higher HBM-bandwidth utilization (Figure~\ref{fig:portability}a) that yields a $1.2$--$2.0\times$ speedup. It also stays high as the architecture scales, exceeding each spec-matched GPU, whereas CUTLASS drops sharply from A100 to GH200 (Figure~\ref{fig:portability}b).

\bibliographystyle{IEEEtran} 
\bibliography{ref}
\end{document}